\documentclass{article}
\usepackage[scale={.83,.83}]{geometry}
\usepackage[latin1]{inputenc}
\usepackage{amsmath,amssymb,bbm,mathrsfs}
\usepackage{hyperref,multicol}

\newlength{\textlength}
\newlength{\overlinelength}
\newcommand{\ovl}[2][.55]{\settowidth{\textlength}{$#2$}
  \setlength{\overlinelength}{3pt}
  \addtolength{\overlinelength}{0.4\textlength}\makebox[\textlength][s]{$#2$}
  \hspace{-.5\textlength}\hspace{-\overlinelength}\hspace{#1\overlinelength}
  \overline{\makebox[\overlinelength][s]{\vphantom{$#2$}}}
  \hspace{-#1\overlinelength}\hspace{.5\textlength}}

\newcommand{\gut}{\text{\sc gut}}
\newcommand{\VEV}[1]{\left\langle #1\right\rangle}

\begin{document}

\begin{center}
  {\Large\bfseries Higher-dimensional operators in SUSY {\sffamily
      SO(10)} GUT models}
  \\[9pt]
  {\large
    J. Sayre and S.~Wiesenfeldt}
  \\[3pt]
  {\normalsize {\slshape
      \begin{minipage}{.8\linewidth}
        \centering Department of Physics, University of Illinois at
        Urbana-Champaign,
        \\
        1110 West Green Street, Urbana, IL 61801, USA
      \end{minipage}
    }}
\end{center}

\begin{abstract}
  \noindent
  {\sffamily SO(10)} GUT models with only small Higgs fields use
  higher-dimensional operators to generate realistic fermion mass
  matrices.  In particular, a Higgs field in the spinor
  representation, $\mathsf{16}^d_H$, acquires a weak scale vev.  We
  include the weak vev of the corresponding field
  $\ovl{\mathsf{16}}^u_H$ and investigate its effects based on two
  successful models, one by Albright and Barr (AB) and another by
  Babu, Pati and Wilczek (BPW).  We find that the BPW model is a
  particular case within a class of models with identical fermion
  masses and mixings.  In contrast, we expect corrections to the
  parameters of AB-type models.
\end{abstract}

\begin{multicols}{2}

  {\slshape Introduction.}
  {\sffamily SO(10)} is the most attractive candidate for a Grand
  Unified Theory (GUT): The matter particles of each standard model
  generation, together with one singlet, the right-handed neutrino,
  are unified in a single representation of a simple gauge group.  The
  breaking of supersymmetric {\sffamily SO(10)} at $M_\gut\simeq
  2\cdot 10^{16}$\,GeV, as suggested by the running of the gauge
  couplings in the MSSM, naturally explains light but massive
  neutrinos.
  
  Over the last several years, driven by the new and more precise
  neutrino data, the search for consistent {\sffamily SO(10)} GUT
  models has intensified \cite{reviews}.  Two models, which use only
  small Higgs representations, are particularly interesting, namely,
  those by Albright and Barr (AB) \cite{ab} and by Babu, Pati and
  Wilczek (BPW) \cite{Babu:1998wi}, with lopsided and hierarchical
  mass matrices, respectively.  This choice of Higgs representations
  guarantees that the theory is perturbative up to the Planck scale.
  Then higher-dimensional operators are included to generate viable
  fermion masses and mixings.\footnote{This is similar to {\sffamily
      SU(5)} where higher-dimensional operators are used for a
    consistent model \cite{Wiesenfeldt:2004qa}.}  These operators
  naturally produce neutrino masses at the observed scale of 0.1\,eV.
  
  The Higgs multiplets used in both models are $\mathsf{10}_H$,
  $\mathsf{16}_H$, $\ovl{\mathsf{16}}_H$ and $\mathsf{45}_H$.
  {\sffamily SO(10)} is broken by the vacuum expectation values (vevs)
  $\VEV{\mathsf{45}_H}\propto\left(B-L\right)$ and
  $\VEV{\mathsf{16}_H}$ and $\VEV{\ovl{\mathsf{16}}_H}$ in the
  {\sffamily SU(5)} singlet direction.  Then the
  $\mathsf{SU(2)_L}$-doublets in $\mathsf{10}_H$ and $\mathsf{16}^d_H$
  acquire electroweak vevs.  (Since the AB model contains two distinct
  fields and the BPW model only one, we use $\mathsf{16}^d_H$ to
  denote the electroweak vev of the $\mathsf{SU(2)_L}$-doublet
  component of $\mathsf{16}_H$ and $\mathsf{16}^\prime_H$ in the BPW
  and AB models, respectively.)  With only $\mathsf{16}_i
  \mathsf{16}_j \mathsf{10}_H$, the fermion mass matrices would
  coincide, $h^u=h^d=h^e=h^\nu$, and the CKM matrix would be unity.
  The dimension-five operator $\mathscr{O}^{B-L} \equiv \mathsf{16}_i
  \mathsf{16}_j \mathsf{10}_H \mathsf{45}_H$ couples differently to
  quarks and leptons, yielding $h^u=h^d\not=h^e=h^\nu$.  Next,
  $\mathscr{O}^d \equiv \mathsf{16}_i \mathsf{16}_j \mathsf{16}_H
  \mathsf{16}^d_H$ contributes to down quarks and charged leptons only
  and enables a non-trivial CKM matrix.  Finally, Majorana neutrino
  masses are generated via couplings \mbox{$\mathscr{O}^N \equiv
    \mathsf{16}_i \mathsf{16}_j \ovl{\mathsf{16}}_H
    \ovl{\mathsf{16}}_H$}.
  
  Both models agree very well with observation and have been studied
  with increasing sophistication \cite{detail}.  Thus one might ask if
  these models are unique or if there are more general classes of
  models that naturally yield the same fermion masses and mixings.  In
  this letter, we consider all dimension-five operators that can
  appear with the present fields, namely $\mathscr{O}^N$,
  $\mathscr{O}^{B-L}$, $\mathscr{O}^d$ and $\mathscr{O}^u \equiv
  \mathsf{16}_i \mathsf{16}_j \ovl{\mathsf{16}}_H
  \ovl{\mathsf{16}}^u_H$ so that the couplings for the Dirac mass
  matrices read
  \begin{multline} 
    h_{ij} \mathsf{16}_i \mathsf{16}_j \mathsf{10}_H + \frac{1}{M}
    \left[ a_{ij} \mathsf{16}_i \mathsf{16}_j \mathsf{10}_H
      \mathsf{45}_H \vphantom{\frac{1}{M}} \right.
    \\[-3pt]
    + \left. g_{ij} \mathsf{16}_i \mathsf{16}_j \mathsf{16}_H
      \mathsf{16}^d_H + g^\prime_{ij} \mathsf{16}_i \mathsf{16}_j
      \ovl{\mathsf{16}}_H \ovl{\mathsf{16}}^u_H \vphantom{\frac{1}{M}}
    \right] .
    \label{eq:couplings}
  \end{multline}
  Similarly to $\mathsf{16}^d_H$, $\ovl{\mathsf{16}}^u_H$ can acquire
  an electroweak vev and then $\mathscr{O}^u$ contributes to up quarks
  and neutrinos.  Whether one or both vevs of $\mathsf{16}^d_H$ and
  $\ovl{\mathsf{16}}^u_H$ are different from zero, depends on the
  particular superpotential.  In the AB model, a non-zero vev for
  $\ovl{\mathsf{16}}^u_H$ is not allowed by the F-flat direction
  constraints on the Higgs superpotential.  In the BPW model,
  $\ovl{\mathsf{16}}^u_H$ can, in general, have a non-vanishing vev;
  however, $\mathscr{O}^u$ is suppressed due to the different charges
  of $\mathsf{16}^d_H$ and $\ovl{\mathsf{16}}^u_H$ under the
  {\sffamily U(1)} family symmetry.  In this letter, we assume that
  with a different choice of family symmetries, both vevs are non-zero
  and $\mathscr{O}^u$ is present.  We use the mass matrices of the AB
  and BPW models to study the effects of a non-vanishing contribution
  from the operator $\mathscr{O}^u$ on fermion masses and mixings.
  For simplicity, we restrict ourselves to just the second and third
  generations.

  \medskip

  {\slshape Hierarchical Models.}
  We start with the case of the BPW model, with only one set of Higgs
  fields in the spinor representation and small contributions from the
  dimension-five operators.  Let us assume that the contributions of
  $\mathscr{O}^{u,d}$ arise from integrating out heavy {\sffamily 10}
  fields [$\mathscr{O}_\mathsf{10}^{u,d}$], as is the case in the AB
  and BPW models; we will discuss the general case later.  We denote
  the couplings as
  \begin{align}
    h_{33} & = 1 \, , & g_{23} \frac{\VEV{\mathsf{16}_H}}{M}\,
    \tan\gamma^d & = \eta \, , & a_{23} \frac{\VEV{\mathsf{45}_H}}{M}
    & = \epsilon \, , \nonumber
    \\[3pt]
    h_{23} & = \sigma \, , & g_{23}^\prime
    \frac{\VEV{\ovl{\mathsf{16}}_H}}{M}\, \tan\gamma^u & = \delta_{10}
    \, ,
  \end{align}
  with
  \begin{align}
    \tan\gamma^d & \equiv
    \frac{\VEV{\mathsf{16}^d_H}}{\VEV{\mathsf{10}^d_H}} \, , &
    \tan\gamma^u & \equiv
    \frac{\VEV{\ovl{\mathsf{16}}^u_H}}{\VEV{\mathsf{10}^u_H}}\, ;
  \end{align}
  the $\left(22\right)$-couplings are assumed to be sufficiently small
  so that they can be ignored.  In the BPW model, $\delta_{10}\equiv0$
  and $\sigma,\epsilon,\eta\sim0.1$.
  
  The mass matrices are given by
  \begin{align}
    h^u & =
    \begin{pmatrix}
      0 & \epsilon+\sigma+\delta^u \cr -\epsilon+\sigma+\delta^u & 1
    \end{pmatrix}
    , \nonumber
    \\[3pt]
    h^d & =
    \begin{pmatrix}
      0 & \epsilon+\sigma+\eta \cr -\epsilon+\sigma+\eta & 1
    \end{pmatrix}
    , \nonumber
    \\[3pt]
    h^\nu & =
    \begin{pmatrix}
      0 & -3\epsilon+\sigma+\delta^\nu \cr 3\epsilon+\sigma+\delta^\nu
      & 1
    \end{pmatrix}
    , \nonumber
    \\[3pt]
    h^e & =
    \begin{pmatrix}
      0 & -3\epsilon+\sigma+\eta \cr 3\epsilon+\sigma+\eta & 1
    \end{pmatrix}
    , \label{eq:masses}
  \end{align}
  where $\delta^u=\delta^\nu=\delta_{10}$.  Eqs.~(\ref{eq:masses}) are
  defined in the basis where the $\mathsf{SU(2)_L}$-doublets are on
  the left and the singlets on the right.  The matrices coincide with
  those of the BPW model with (see Eq.~(15) in
  Ref.~\cite{Babu:1998wi})
  \begin{align} \label{eq:compare-bpw}
    \epsilon_\text{\sc bpw} & = \epsilon \, , & \sigma_\text{\sc bpw}
    & = \sigma + \delta^u \, , & \eta_\text{\sc bpw} & = \sigma+\eta
    \, .
  \end{align}
  Thus any realization within this class of models can reproduce the
  fermion mass matrices of the BPW model.  In principle, one could set
  any one of $\sigma$, $\eta$ (i.e. the operator $\mathscr{O}^d$) or
  $\delta^u$ (i.e. $\mathscr{O}^u$) to zero.
  
  In Eqs.~(\ref{eq:masses}), we assumed that $\mathscr{O}^{u,d}$ was
  obtained using heavy {\sffamily 10} fields.  Unlike $\mathscr{O}^d$,
  the operator $\mathscr{O}^u$ can arise from integrating out heavy
  singlets or fields in the adjoint representation as well.
  Integrating out heavy singlet fields [$\mathscr{O}_\mathsf{1}^u$]
  gives only terms to $h^\nu$, and $\mathscr{O}_\mathsf{45}^u$
  contributes to $h^u$ and $h^\nu$ in the ratio $8:3$ \cite{o^u}.
  Since $\delta^u\not=\delta^\nu$, we expect this more general class
  of models to differ from the BPW model but the differences appear in
  $h^\nu$ and affect the neutrino masses and mixings only.  The
  atmospheric mixing angle is given by
  \begin{align}
    \tan\theta^\nu_{\mu\tau} & =
    \sqrt{\left|\frac{m_{\nu_2}}{m_{\nu_3}}\right|} \simeq
    \frac{h^N_{23} \left(3\epsilon-\sigma-\delta^\nu\right)}{h^N_{33}
      \left(3\epsilon+\sigma+\delta^\nu\right) - 2 h^N_{23}} \, ,
  \end{align}
  where $h^N_{23}\sim\delta_1+\delta_{45}$ and
  $\frac{h^N_{23}}{h^N_{33}}\equiv y_\text{\sc bpw}$ (see Eq.~(22) in
  Ref.~\cite{Babu:1998wi}).  In the BPW model, $\delta^\nu=\delta^u$
  and $h^N_{33}=1$.
  
  Finally, let us remark that for two distinct fields,
  $\ovl{\mathsf{16}}_H$ and $\ovl{\mathsf{16}}^u_H$, the contribution
  to $h^\nu$ is not symmetric, since the neutrino singlet couples to
  $\ovl{\mathsf{16}}_H$ and the lepton doublet to
  $\ovl{\mathsf{16}}^u_H$.  Additionally, if we allow heavy {\sffamily
    120} fields, the contributions to $\mathscr{O}_\mathsf{120}^{u,d}$
  are not symmetric either.

  \medskip

  {\slshape Lopsided Models.}
  Let us turn to models where the (23) and (32)-elements of the mass
  matrices are not of the same order.  In the AB model with two
  distinct pairs of Higgs fields in the spinor representation, the
  $\mathsf{U(1)} \times \mathbbm{Z}_2 \times \mathbbm{Z}_2$ family
  symmetry allows only $\left[ \mathsf{16}_2 \mathsf{16}^d_H
  \right]_\mathsf{10} \left[ \mathsf{16}_3 \mathsf{16}_H
  \right]_\mathsf{10}$.  In addition, $\sigma\equiv0$ (see
  Eqs.~(\ref{eq:masses})).  As a result, $h^d$ and $h^e$ have lopsided
  structures, and a large value for $\tan\gamma^d$ leads to
  $\eta\simeq 1.8$, whereas $\epsilon\simeq0.05$ \cite{ab}.
  
  Let us assume that a specific family symmetry leads to $h^d$ and
  $h^e$ as in the AB model and that $\mathscr{O}^u$ contributes to
  $h^u$ and $h^\nu$.  Then the quark matrices read\footnote{In order
    to have a consistent notation, we replace $\epsilon/3$ and
    $\sigma$ in Refs.~\cite{ab} by $-\epsilon$ and $\eta$; furthermore
    the mass matrices in Refs.~\cite{ab} are defined in the basis with
    the singlets on the left and the doublets on the right.}
  \begin{align} \label{eq:lopsided}
    h^u & =
    \begin{pmatrix} 
      0 & \epsilon+\delta^u \cr -\epsilon+\delta^u & 1
    \end{pmatrix}
    , & h^d & =
    \begin{pmatrix} 
      0 & \epsilon \cr -\epsilon+\eta & 1
    \end{pmatrix}
    ,
  \end{align}
  with $\delta^u=\delta_{10}+\delta_{45}$.  In the AB model,
  $\delta^u\equiv0$.  Note that the contribution from $\mathscr{O}^u$
  to $h^u$ is symmetric, regardless of the family symmetry, because
  the quark doublet and the up-quark singlet belong to the same
  {\sffamily SU(5)}-representation.  In contrast, the down quark and
  lepton doublets are in different {\sffamily SU(5)}-representations
  than the singlets.
  
  To obtain a viable hierarchy for the up quarks, $\delta^u$ must be
  small.  Then we obtain
  \begin{align} \label{eq:lopsided-ev}
    \frac{m_c}{m_t} \simeq \left|\delta^2-\epsilon^2\right| ,
    \mspace{9mu} %
    \frac{m_s}{m_b} \simeq \frac{\epsilon\eta}{1+\eta^2} \, ,
    \mspace{9mu} %
    V_{cb} \simeq \frac{\epsilon\eta^2}{1+\eta^2}+\delta \, .
  \end{align}
  Unlike in the BPW case, we have found an additional parameter in
  $h^u$, which cannot be absorbed in the original parameters and needs
  to be included in the fit.  On the other hand, this parameter has to
  be small in order to reproduce the quark masses and mixing.  We do
  not expect significant changes to the neutrino sector, in particular
  the mixing angle is still dominated by the lopsidedness of $h^e$.
  
  For the hierarchical models, we found that we could set $\eta=0$ and
  still reproduce the results of the BPW model with $\delta^u\equiv0$
  (cf.  Eq.~(\ref{eq:compare-bpw})).  Thus one might consider the case
  $\eta=0$ for the AB-type models with mass matrices as displayed in
  Eqs.~(\ref{eq:lopsided}); however, they rely on the lopsidedness of
  $h^d$ and $h^e$.  Hence, one could try to mimic these effects with a
  lopsided matrix $h^u$.
  
  A lopsided structure for $h^u$ can be generated if we take into
  account antisymmetric contributions and tune them such that
  different terms in the mass matrices cancel.  We already know that
  $\mathscr{O}^{B-L}$ gives antisymmetric contributions (see
  Eqs.~(\ref{eq:masses})) but $\epsilon\sim0.1$ is too small to
  generate lopsided entries.  As already mentioned, we can also
  integrate out heavy fields in the (large) {\sffamily 120}
  representation, and $\mathscr{O}^u_\mathsf{120}=\left[ \mathsf{16}_i
    \mathsf{16}_j \right]_\mathsf{120} \left[ \ovl{\mathsf{16}}_H
    \ovl{\mathsf{16}}^u_H \right]_\mathsf{120}$ yields an
  antisymmetric contribution, $\delta_{120}$.  If
  $\delta_{120}=-\delta$, then
  \begin{align}
    h^u & = 
    \begin{pmatrix}
      0 & \epsilon \cr -\epsilon+2\delta & 1
    \end{pmatrix}
    ,
  \end{align}
  and we get
  \begin{align}
    \frac{m_c}{m_t} & \simeq \frac{2\epsilon\delta}{1+4\delta^2} \, ,
    & V_{cb} & \simeq \epsilon\left(\frac{1}{1+4\delta^2}
      -\frac{1}{1+\eta^2}\right) .
  \end{align}
  Compared to Eqs.~(\ref{eq:lopsided}), the factor
  $\frac{2\delta}{1+4\delta^2}$ must be of order $\epsilon$ in order
  to enable the up quark mass hierarchy.  The two solutions are
  $\delta\sim\epsilon$ and $\delta\sim\frac{1}{\epsilon}$, but if
  $\eta=0$, we obtain
  $V_{cb}\simeq\frac{4\epsilon\delta^2}{1+4\delta^2}$ so that
  $\delta\simeq1$.  Hence, $\eta=0$ is not a viable case.
  
  With both $\delta$ and $\eta$ non-zero, one should consider what the
  natural scale for $\delta$ is.  Since
  $\frac{\VEV{\mathsf{10}^u_H}}{\VEV{\mathsf{10}^d_H}}$ can be much
  bigger than one, a large value for $\tan\gamma^d$ is not unnatural
  \cite{ab}.  On the other hand, we expect $\tan\gamma^u$ not to be
  bigger than one so that $\delta\sim\epsilon\sim0.1$ is a natural
  choice.  Therefore we do not consider solutions with
  $\delta\sim\frac{1}{\epsilon}\gg1$.

  \medskip

  {\slshape Conclusions.}
  We have shown that the BPW model belongs to a class of models with
  equivalent fermion masses and mixings; it represents the special
  case where the operator $\mathscr{O}^u$ vanishes.  If we factor in
  heavy singlets and heavy fields in the adjoint representation, then
  $\mathscr{O}^u$ does not contribute equally to $h_u$ and $h_\nu$.
  Therefore we expect deviations from the BPW model in the neutrino
  sector, which are presumably small.  In AB-type models, including
  $\mathscr{O}^u$ generically introduces a new parameter in the quark
  mass matrices which is smaller than the contribution from the
  operator $\mathscr{O}^d$.  Unlike the down quarks and leptons,
  family symmetries cannot generate a lopsided up-quark mass matrix
  unless we allow for fine-tuning.
  
  We restricted the discussion to the second and third generations
  only; however, the results apply to the three-generational case as
  well.

  \smallskip
  
  We are grateful for valuable discussions with S.~Willenbrock and for
  useful comments on the manuscript by C.~Albright and J.~Pati.
  This work was supported in part by the U.~S.~Department of Energy
  under contract No.~DE-FG02-91ER40677.


{\small
}

\end{multicols}


\begin{thebibliography}{9}
    \addtolength{\itemsep}{-3pt}
  \bibitem{reviews} For recent reviews on {\sffamily SO(10)} models,
    see
    J.~C.~Pati, hep-ph/0507307;
    A.~Melfo and G.~Senjanovic, hep-ph/0511011.

  \bibitem{ab}
    C.~H.~Albright and S.~M.~Barr, Phys.\ Rev.\ D {\bf 58}, 013002
    (1998);
    C.~H.~Albright, K.~S.~Babu and S.~M.~Barr, Phys.\ Rev.\ Lett.\ 
    {\bf 81}, 1167 (1998);
    C.~H.~Albright and S.~M.~Barr, Phys.\ Lett.\ B {\bf 452}, 287
    (1999);
    \emph{ibid}\ {\bf 461}, 218 (1999);
    Phys.\ Rev.\ Lett.\ {\bf 85}, 244 (2000);
    Phys.\ Rev.\ D {\bf 62}, 093008 (2000).
    
  \bibitem{Babu:1998wi} K.~S.~Babu, J.~C.~Pati and F.~Wilczek,
    Nucl.\ Phys.\ B {\bf 566}, 33 (2000).

  \bibitem{Wiesenfeldt:2004qa}
    S.~Wiesenfeldt,
    Mod.\ Phys.\ Lett.\ A {\bf 19}, 2155 (2004)
    and references therein.
  
  \bibitem{detail}
    C.~H.~Albright and S.~M.~Barr, Phys.\ Rev.\ D {\bf 64}, 073010
    (2001);
    \emph{ibid}\ {\bf 69}, 073010 (2004);
    C.~H.~Albright,
    Phys.\ Rev.\ D {\bf 72}, 013001 (2005);
    K.~S.~Babu, J.~C.~Pati and P.~Rastogi, Phys.\ Rev.\ D {\bf 71},
    015005 (2005);
    Phys.\ Lett.\ B {\bf 621}, 160 (2005);
    P.~Rastogi, Phys.\ Rev.\ D {\bf 72}, 075002 (2005).
    
  \bibitem{o^u}
    K.~S.~Babu and S.~M.~Barr, Phys.\ Rev.\ Lett.\ {\bf 85}, 1170
    (2000);
    P.~Nath and R.~M.~Syed, Nucl.\ Phys.\ B {\bf 618}, 138 (2001).

  \end{thebibliography}
\end{document}